\documentclass[english,aps,preprint, nofootinbib,superscriptaddress,11pt]{article}
\pdfoutput=1
\usepackage[T1]{fontenc}
\usepackage{geometry}
\usepackage[latin9]{inputenc}
\usepackage{amsmath,amssymb}
\usepackage{graphicx}

\makeatletter
\@ifundefined{textcolor}{}
{%
 \definecolor{BLACK}{gray}{0}
 \definecolor{WHITE}{gray}{1}
 \definecolor{RED}{rgb}{1,0,0}
 \definecolor{GREEN}{rgb}{0,1,0}
 \definecolor{BLUE}{rgb}{0,0,1}
 \definecolor{CYAN}{cmyk}{1,0,0,0}
 \definecolor{MAGENTA}{cmyk}{0,1,0,0}
 \definecolor{YELLOW}{cmyk}{0,0,1,0}
 }

\makeatother

\usepackage{babel}

\title{
\vspace*{-2cm}
\begin{flushright}
\normalsize{EFI-11-11 \\
ANL-HEP-PR-11-28\\ 
FERMILAB-PUB-11-189-T\\}
~\\
\end{flushright}
\vspace*{1.5cm}
$W$ plus two jets from a quasi-inert Higgs doublet \\
\author{\textbf{Qing-Hong Cao$^{a,b}$, Marcela Carena$^{b,c}$, Stefania Gori$^b$, Arjun Menon$^d$, }\\
\textbf{Pedro  Schwaller$^{a,e}$, Carlos E.M. Wagner$^{a,b,f}$ and Lian-Tao Wang$^{b,f}$} \\
~\\
\normalsize\emph{$^a$HEP Division, Argonne National Laboratory, 9700 Cass Ave., Argonne, IL 60439}\\
\normalsize\emph{$^b$Enrico Fermi Institute, University of Chicago, Chicago, IL 60637} \\
\normalsize\emph{$^c$Theoretical  Physics Department, Fermilab, Batavia, IL 60510}\\
\normalsize\emph{$^d$Illinois Institute of Technology, Chicago, IL 60616}\\
\normalsize\emph{$^e$ Physics Department, University of Illinois at Chicago, Chicago, IL 60607}\\
\normalsize\emph{$^f$Kavli Institute for Cosmological Physics, University of Chicago, Chicago, IL 60637}
}}

\begin{document}

\maketitle
\vspace{1cm}

\begin{abstract}
We show that, the  result recently reported  by the CDF collaboration showing  an excess in the invariant mass distribution of jet pairs produced in association with a  $W$-boson can be explained by a simple extension of the Standard Model (SM) with an additional quasi-inert Higgs doublet. The two additional neutral Higgs states $H^0$ and $A^0$ have a mass of about 150~GeV and decay into a pair of jets. $W^\pm H^0/A^0$ pairs are produced from the decay of the heavier charged Higgs boson $H^\pm$.  
Depending on the precise masses of the neutral and charged Higgs bosons, the model   is shown to be in agreement with constraints from electroweak precision tests and from flavor physics
for a broad range of the Standard Model-like Higgs mass from 100 GeV to several hundreds of GeV.  Other possible signals of this model at the Tevatron and the LHC are discussed. 
\end{abstract}
\thispagestyle{empty}
\newpage

\section{Introduction}

Recently the CDF collaboration has reported an excess in the dijet invariant mass distribution in events where a $W$-boson is produced together with two jets. The $W$-boson in this case is required to decay leptonically into electrons or muons. Besides the peak around $M_{jj} = 80$~GeV from di-boson production with one hadronically decaying $W$-boson, the invariant mass  distribution shows an excess in the region $120$~GeV$<M_{jj}<160$~GeV. This  disagrees with the Standard Model (SM) background at the $3.2~\sigma$ level and is consistent with a dijet resonance with a mass of $144\pm 5$~GeV~\cite{Aaltonen:2011mk}. 

The CDF excess can readily be explained by introducing an additional particle that couples to the first generation of quarks, e.g. using a $Z'$ boson~\cite{Buckley:2011vc,Yu:2011cw,Cheung:2011zt,Anchordoqui:2011ag,Buckley:2011vs,Ko:2011ns,Fox:2011qd,Jung:2011ue}. In this case one has to worry about a potentially large contribution also in the $Z$+2jets and $\gamma$+2jets channels, which are only suppressed by electroweak mixing angles and by the branching fraction of the $Z$ boson into leptons. 

Another promising approach is to introduce a second resonance that is produced on-shell and decays into a $W$-boson and into the dijet resonance. For recent work that implements this and other ideas, see~\cite{Eichten:2011sh,Kilic:2011sr,Wang:2011uq,Nelson:2011us,Sato:2011ui,Wang:2011ta,He:2011ss,Dobrescu:2011px}. It is also possible that the excess is due to mis-modeling  of some of the Standard Model backgrounds, e.g. from single top production~\cite{Sullivan:2011hu,Plehn:2011nx}.

In this paper, we present a simple and renormalizable model that implements the two resonance approach in a compact way. In addition to the Standard Model particle content, just one complex electroweak scalar doublet is introduced. In the next section, we introduce the basic features of the model. In section 3 we discuss constraints from electroweak precision tests and from flavor physics. In section 4 we show that the model can explain the excess reported by CDF, and, in section 5, we discuss possible other signals at the Tevatron and at the LHC. We reserve section 6 for our conclusions and we present an Appendix, in which we analyze the possible consequences of a different mass arrangement for the charged and neutral Higgs bosons of the theory.

\section{The model}

We consider a simple extension of the Standard Model by including an additional
scalar doublet $\Phi_{2}$, with the same hypercharge as the SM doublet $\Phi_{1}$. Consider
the renormalizable scalar potential of $\Phi_{1}=(\phi^{-},\phi^{0})$ and $\Phi_{2}=[H^{-},(H^{0}+iA^{0})/\sqrt{2}]$,
\begin{eqnarray}\label{eq:potential}
V & = & \mu_{1}^{2}\Phi_{1}^{\dagger}\Phi_{1}+\mu_{2}^{2}\Phi_{2}^{\dagger}\Phi_{2}+\left(\mu_{12}^2 \Phi_1^\dagger \Phi_2 + {\rm h.c.}\right)+\frac{1}{2}\lambda_{1}\left(\Phi_{1}^{\dagger}\Phi_{1}\right)^{2}+\frac{1}{2}\lambda_{2}\left(\Phi_{2}^{\dagger}\Phi_{2}\right)^{2}\notag\\
 &  &+\lambda_{3}\left(\Phi_{1}^{\dagger}\Phi_{1}\right)\left(\Phi_{2}^{\dagger}\Phi_{2}\right) +\lambda_{4}\left(\Phi_{1}^{\dagger}\Phi_{2}\right)\left(\Phi_{2}^{\dagger}\Phi_{1}\right)+\frac{1}{2}\lambda_{5}\left(\Phi_{1}^{\dagger}\Phi_{2}\right)^{2}+\frac{1}{2}\lambda_{5}^{*}\left(\Phi_{2}^{\dagger}\Phi_{1}\right)^{2},
\end{eqnarray}
where we have assumed that all quartic couplings involving an odd number of $\Phi_1$ or $\Phi_2$ fields are suppressed (see discussion below).
 All the above parameters are necessarily real except $\lambda_{5}$ and $\mu_{12}^2$. We note also that vacuum stability imposes the requirements 
\begin{equation}
\lambda_{1,2}>0\,,\qquad\lambda_{3} > - \sqrt{\lambda_1 \lambda_2}\,, \qquad \lambda_{3}+\lambda_{4}-\left|\lambda_{5}\right|>-\sqrt{\lambda_{1}\lambda_{2}}\,.\label{eq:vacuumstability}
\end{equation}
These conditions are easily fulfilled for the range of parameters relevant to this work. 

We assume that $\Phi_{1}$ and $\Phi_{2}$ interact with the SM quarks via
the Yukawa interaction
\begin{align}\label{eq:yukawaL}
\mathcal{L}\supset y_{u,1}\Phi_{1}Qu+y_{d,1}\Phi_{1}^{c}Qd+y_{u,2}\Phi_{2}Qu+y_{d,2}\Phi_{2}^{c}Qd\,,
\end{align}
where $Q=(u_{L},d_{L})^{c}$ denotes the SM quark doublet while $u$
and $d$ are the SM quark singlets. To avoid experimental constraints from the lepton sector, at this point we choose not to couple $\Phi_2$ to leptons.

\bigskip

We demand that $\Phi_{2}$ develops a vanishing (or negligible) vacuum expectation value (VEV). In
other words $\Phi_{2}$ is not involved in the electroweak symmetry breaking. 
The SM $SU(2)\times U(1)_{Y}$ gauge symmetry is spontaneously
broken by $\left\langle \phi^{0}\right\rangle =v=174\,{\rm GeV}$.
In order to forbid a VEV for $\Phi_{2}$ at tree-level it suffices to set  
 the $\mu_{12}$ term in the scalar potential to zero. At one loop however, the $\mu_{12}$ term will be generated by radiative corrections:

\begin{align}
	\delta \mu_{12}^2 = \frac{y_{u,1}^\dagger y_{u,2}+y_{d,1}^\dagger y_{d,2}}{16 \pi^2} \Lambda^2\,,
\end{align}
where $\Lambda$ is a cutoff scale that is naturally expected to be around a few TeV. Assuming that the $\Phi_2$ couplings are flavor independent, and of order 0.1, we get a mixing contribution of  the order of 1000~GeV$^2$.  This mixing term would induce a vacuum expectation value $\langle \Phi_2\rangle$ of order 10~GeV, leading to too large contributions of $\Phi_2$ to the light quark masses. In order to avoid such a large contribution, a mixing mass $\mu_{12}^2$ smaller than about  $10$~GeV$^2$ would be required, what may be obtained in the case of flavor independent couplings by a cancellation between the tree-level  and loop-induced contributions, requiring a large fine tuning.   It is however possible that  $\Phi_2$ only couples to the first generation quarks.  In such case, provided that $\mu_{12}^2$ is set to zero at the tree-level by some symmetry, no large fine tuning would be required.

To explicitly forbid the mixing between the two Higgs doublets, we can impose a $\mathbf{Z}_2$ parity symmetry under which $\Phi_2$ is odd~\cite{Ma:2006km,LopezHonorez:2006gr}, leading also to a justification of the suppression of the quartic couplings with odd powers of the fields $\Phi_{1,2}$ in the scalar potential (Eq. (\ref{eq:potential})).  This is similar to what is obtained in R-Parity violating models, with $\Phi_2$ replaced by a slepton~\cite{Kilic:2011sr}. However, R-Parity violating models of this type are strongly constrained by the possible generation of neutrino masses (for a recent analysis, see Ref.~\cite{Dreiner:2010ye}), a constraint that is not present in the two Higgs doublet model case.

To allow a coupling of $\Phi_2$ to quarks, we further require that the right-handed up quark is odd under this new parity symmetry, while all the other SM fields are even. A consequence of this is that $\Phi_1$ would not couple to the right-handed up quark and therefore the up quark would remain massless at tree-level.  Were this parity symmetry preserved, it would lead to a potential solution of the strong CP-problem, which seems to be disfavored by results from chiral perturbation and lattice gauge theory~\cite{Leutwyler:1989pn,Aubin:2004fs,Nakamura:2010zzi} (see, however, Ref.~\cite{Davoudiasl:2007zx}).   

In order to generate a value of the up quark mass of about a few MeV, a soft breakdown of the $Z_2$-symmetry, via a non-vanishing $\mu_{12}^2$ term 
of about 10~GeV$^2$, would be necessary.

\bigskip

Assuming this approximate $Z_2$-symmetry in the Yukawa Lagrangian of Eq. (\ref{eq:yukawaL}), we can write down the Yukawa couplings that we use for our analysis as
\begin{equation}
\mathcal{L}_{\mathcal{Y}}=y_{u,1}^{i\alpha}\Phi_{1}Q_iu_\alpha+y_{d,1}^{ij}\Phi_{1}^{c}Q_i d_j+y^{i1}_{u,2}\Phi_{2}Q_i u_1\,.
\end{equation}
Note that the index $\alpha$ can only take the values $\alpha\in \{ 2,3\}$ since a coupling of the right-handed up quark to $\Phi_1$ is forbidden by the parity symmetry. 

We conclude this section with the Higgs spectrum of the theory:
\begin{eqnarray}
M_h^2 & = & 2\lambda_{1}v^{2}\,,\\
M_{H^\pm}^2 & = & \mu_{2}^{2}+\lambda_{3}v^{2}\,,\\
M_{H^0}^2 & = & \mu_{2}^{2}+(\lambda_{3}+\lambda_{4}+\lambda_{5})v^{2}\,,\\
M_{A^0}^2 & = & \mu_{2}^{2}+(\lambda_{3}+\lambda_{4}-\lambda_{5})v^{2}\,.
\end{eqnarray}
The lone Higgs boson of the Standard Model is of course $h$, whereas
$H^{\pm}$, $H^{0}$, and $A^{0}$ are the components of the new scalar
doublet.

Using these expressions for the masses of the scalar fields, we can rewrite the condition on the vacuum stability (Eq. (\ref{eq:vacuumstability})) as a condition on the $\mu_2$ parameter 
\begin{equation}
\mu_{2}^{2}<  {\rm Min}(M_{H^{0}}^{2},M_{A^0}^2)+\sqrt{\frac{\lambda_{2}}{2}}M_{h}v\,.\label{eq:vacuumbound}
\end{equation}

Finally, the $Z$-pole precision measurement sets the limit 
\begin{equation}
M_{Z} < M_{H^{0}}+M_{A^{0}}.
\label{boundAH}
\end{equation}
There are also bounds coming from the direct searches for the neutral Higgs particles at LEP2, which may increase 
the above limit in Eq.~(\ref{boundAH}) from $M_Z$ to values 
close to 200 GeV. However, the precise bound on the sum of the neutral Higgs boson masses depends strongly 
on the decay properties of these particles.  In this article, we will be interested in masses for these particles 
that strongly exceed the precision measurement and LEP2 bounds.

\section{Electroweak precision constraints and flavor}\label{Sec:EWPTandflavor}

We perform the analysis of the ElectroWeak Precision Tests (EWPTs) in the S-T plane, with the experimental contours
taken from~\cite{Plot}. Since, as we will show in the next section, we have to require a rather large mass splitting between the scalars $H^0,A^0$ and the charged Higgs $H^\pm$ to fit the $W+2$jets CDF excess, we expect a large positive New Physics (NP) contribution to the T parameter. To calculate the NP effects to S and T, we use the general formula valid for 2HDMs~\cite{Chankowski:1999ta,Chankowski:2000an}, under the assumptions of no mixing between the two Higgs doublets and of negligible VEV of the second doublet $\Phi_2$.  In order to generate a $W + 2j$ signature consistent with observation, we shall fix the mass of one of the scalars to be close to 150~GeV.

\begin{figure}[t]
\center
\includegraphics[width=.6\textwidth]{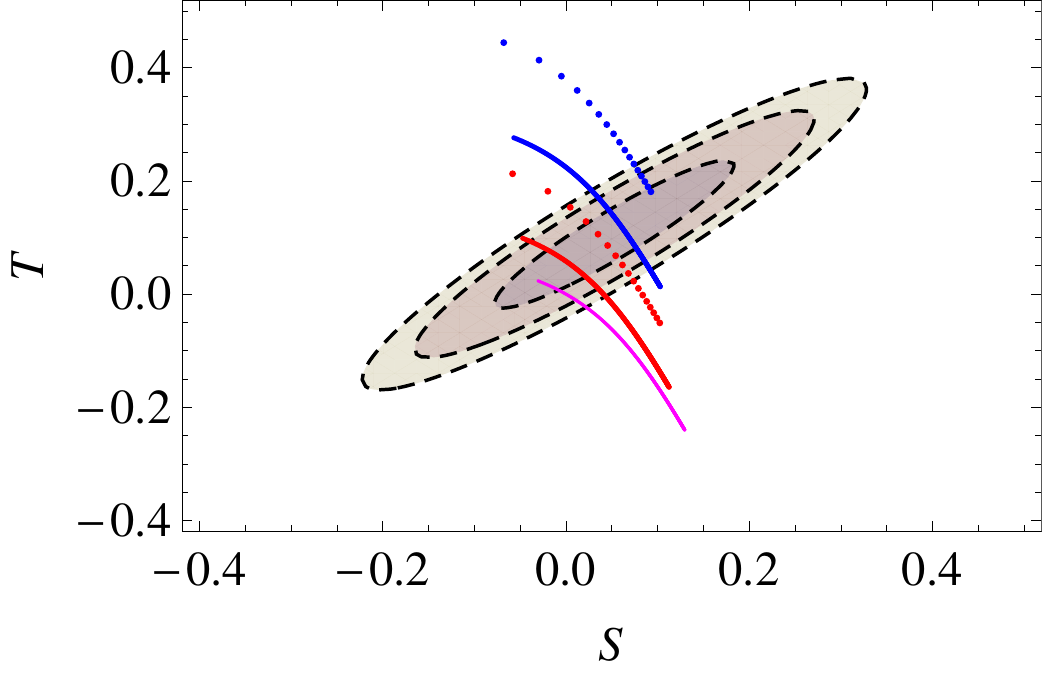}
\caption{The dashed ellipses show the $68\%$, $95\%$ and $99\%$ CL contours on the (S, T) plane
obtained by Gfitter~\cite{Plot}. We fix $M_{H^0}=150$ GeV, $M_{A^0}=210$ GeV and the charged Higgs mass to $M_{H^\pm}=300$ GeV (blue curve, second from top), $M_{H^\pm}=250$ GeV (red curve, fourth from top).  Each  line shows, from top to bottom,  the shift in (S, T) resulting from increasing the SM-like Higgs mass from 80 GeV to 900 GeV.  The case of non splitting between $M_{A^0}$ and $M_{H^0}$ is also shown ($M_{H^0}=M_{A^0}=150$ GeV): the blue dotted curve (first from top) is for $M_{H^\pm}=300$ GeV, the red dotted curve (third from  top) for $M_{H^\pm}=250$ GeV.  The magenta curve (last from top) corresponds to the SM.}
\label{fig:EWPT}
\end{figure}

In Fig.~\ref{fig:EWPT} we present the values of the $S$ and $T$ parameters for varying values of the  SM-like  Higgs mass $M_h$, from 80~GeV to 900~GeV, and  different values of the $\Phi_2$ charged and neutral Higgs bosons masses.  Dashed ellipses represent the values of the S and T parameters consistent with electroweak precision tests at 68, 95 and 99 $\%$ confidence level.   Fixing a splitting of 150 GeV between the charged Higgs and the neutral Higgs bosons ($M_{H^0}=150$ GeV, $M_{H^\pm}=300$ GeV) and assuming degeneracy between scalar and pseudoscalar masses, a very large mass for the SM Higgs boson $h$, $M_h > 650$~GeV, would be required to obtain consistency with electroweak precision tests at the 2$\sigma$ level (see the blue dotted curve in Fig.~\ref{fig:EWPT}). A good fit, for moderate values of the SM Higgs boson mass $M_h$ may be obtained by lowering the mass of the charged Higgs: for  $M_{H^\pm}=250$ GeV and $M_{H^0}=M_{A^0}=150$ GeV, a SM Higgs with mass in the range $(200-480)$ GeV (and with central value 300 GeV) would be required to be in agreement with EWPTs at the $2\sigma$ level (see the red dotted curve in Fig.~\ref{fig:EWPT}).

An alternative way of reducing the required SM-like Higgs boson mass is to assume a splitting between the $A^0$ and $H^0$ masses
through a non-vanishing value of $\lambda_5$\footnote{Note however that the constraints coming from flavor physics are in general harder to satisfy in this case, as will be discussed in the following.}.
In Fig.~\ref{fig:EWPT}, we show our results once we fix $M_{H^0}=150$ GeV, $M_{A^0}=210$ GeV and the charged mass to $M_{H^\pm}=300$ GeV (solid blue curve) or $M_{H^\pm}=250$ GeV (solid red curve).  As we can observe from the figure, a more moderate value of the Higgs mass $M_h$ is required in order to be in accordance with EWPTs: the best fit is obtained for $M_h=450$ GeV (with 2$\sigma$ boundary given by $(290-660)$ GeV) in the case of a 300 GeV charged Higgs boson, and for $M_h=150$ GeV (with 2$\sigma$  boundary given by $(90-250)$ GeV) in the case of $M_{H^\pm}=250$ GeV.  Observe that a  heavy SM-like Higgs boson with mass up to 400~GeV will be tested at the LHC through its decay into massive gauge bosons within the next two years~\cite{CMSHiggs}. 

\bigskip

\begin{figure}[t]
\center
\includegraphics[width=.45\textwidth]{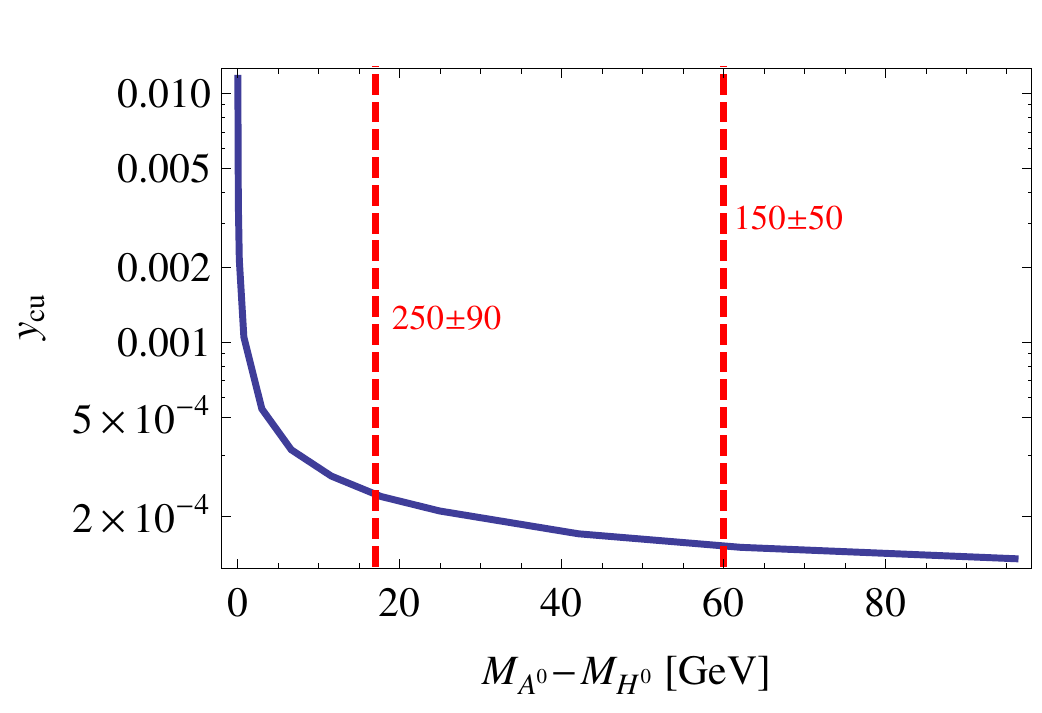}\hspace{0.6cm}\includegraphics[width=.45\textwidth]{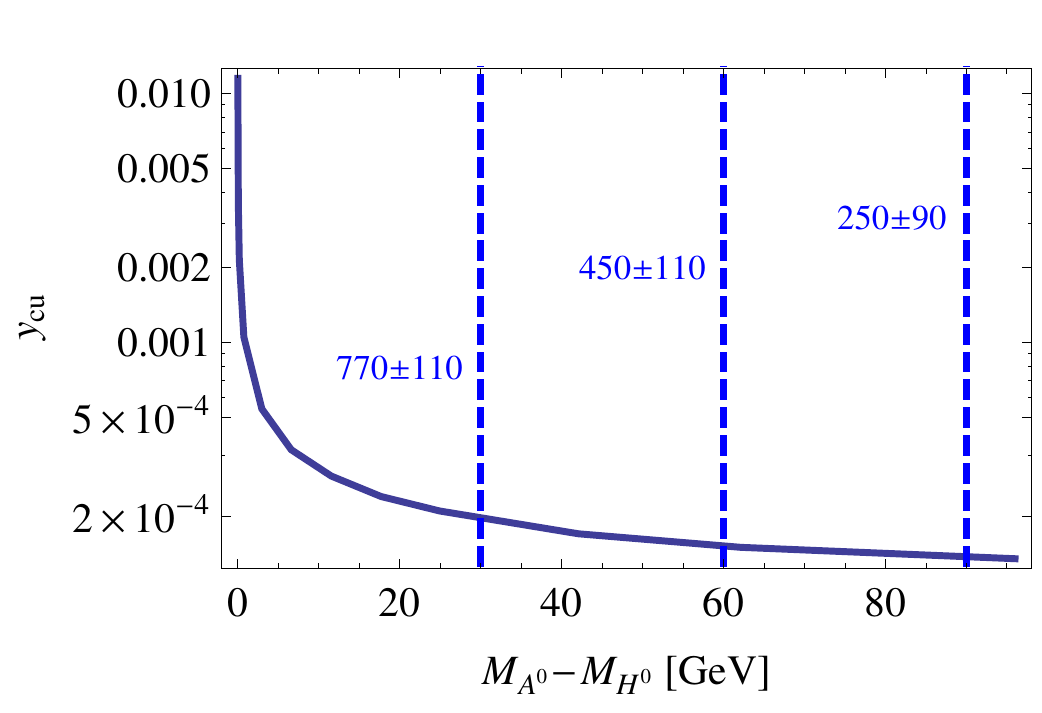}
\caption{Constraint on the coupling $y_{cu}$ as a function of the mass splitting $M_{A^0}-M_{H^0}$. The region above the solid line is
disfavored at 2$\sigma$ by the present experimental bounds on $D-\bar D$ mixing. On the left panel the charged mass is fixed to 250 GeV, on the right $M_{H^\pm}=300$ GeV. The vertical lines indicate the value of the SM Higgs mass $M_h$ (in GeV) needed in order to satisfy the constraints coming from EWPTs at the $1\sigma$ level.}
\label{fig:DD}
\end{figure}

Furthermore flavor adds additional constraints on this model. In the mass eigenbasis,
the quark-$\Phi_2$ Higgs interaction Lagrangian involving the right-handed up quark is
\begin{align}
\mathcal{L}^{u_1} = y_{u,2}^{i1} U_{u,L}^{ij} \left(\phi_2^0  \bar{u}_{L,j} 
u_{1} - \phi_2^- V_{\rm CKM}^{jk} \bar{d}_{L,k} u_{1} \right).
\end{align}
Clearly the GIM mechanism is violated strongly and an assumption
about the flavor structure of the Yukawa matrix is needed to avoid large NP effects in
flavor violating observables. We choose to make the simplifying assumption,
$y_{u,2}^{i1} = y_{u,2} \delta^{i1}$. We shall analyze two possible limits for the
rotation matrix, $U_{u,L}^{ij} = \delta^{ij}$ and $U_{u,L}^{ij} = 
(V_{\rm CKM}^{ij})^\dagger$. Under these assumptions the flavor violating Higgs mediated
couplings are
\begin{align}
\mathcal{L}^{u_1} = \left\{\begin{array}{cc}
y_{u,2} \left(\phi_2^0  \bar{u}_{L,1} u_{1} - 
\phi_2^- V_{\rm CKM}^{1k} \bar{d}_{L,k} u_{1} \right) &{\rm{for}}\,\, U_{u,L}^{ij} = \delta^{ij}\,, \\
y_{u,2} \left(V_{\rm CKM}^{*j1} \phi_2^0 \bar{u}_{L,j} u_{1} - 
\phi_2^- \bar{d}_{L,1} u_{1} \right) &\,\,\,\,\,\,\,\,{\rm{for}}\,\, U_{u,L}^{ij} = V_{\rm CKM}^{*ji}\,. \\
\end{array}\right.
\end{align}

For the case $U_{u,L}^{ij} = \delta^{ij}$, the charged Higgs $\phi_2^-$ interactions
lead to a new source of flavor violation, beyond the SM one.   These flavor violating
interactions are governed by the couplings $y_{u,2} V_{\rm CKM}^{1k}$, and depending
on the $y_{u,2}$ value,  tree-level flavor violating 
processes like $K^+ \to \pi^+ \pi^0$ can put strong constraints on this 
scenario. The contribution to the $K^+ \to \pi^+ \pi^0$ mode due to the charged 
Higgs, as compared to the Standard Model, is naively suppressed by the factor
$(y_{u,2}/g_2)^4 (m_{K}/m_\pi)^2 (M_W/M_{H^\pm})^4$. 
Additionally, NP contributions to Kaon mixing appear only at the loop-level. The
constraints from Kaon physics are satisfied for values of the Yukawa 
couplings smaller than about 0.1 and charged Higgs masses larger than  250~GeV,
as required to describe the $W+$ 2jets data within this model. 

For the case of $U_{u,L}^{ij} = V_{\rm CKM}^{*ji}$, the flavor violating coupling
of the c- and u-quarks to $\phi_2^0$ leads to a generically large tree-level
contribution to the $D-\bar{D}$ mixing when the scalar 
and pseudoscalar masses are non-degenerate.
In all generality, we can define $y_{cu}$ the coupling $\phi_2^0 \bar{u}_{L,2} u_{1}$. 
In Fig.~\ref{fig:DD}, we show the constraint on
$y_{cu}$ as a function of the mass splitting $M_{A^0}-M_{H^0}$. The fit is 
performed using the experimental values given by HFAG~\cite{HFAG}. As 
Fig.~\ref{fig:DD} suggests, even a few~GeV splitting between $M_{H^0}$ and 
$M_{A^0}$ requires that $y_{cu} \lesssim 10^{-3}$. This bound can be read as a bound on the 
coupling $y_{u,2}$ for the scenario with $U_{u,L}^{ij} = V_{\rm CKM}^{*ji}$ ($y_{cu}=y_{u,2}V_{\rm CKM}^{*21}$).
We can conclude that for this scenario to be viable we need $H^0$ and $A^0$ to be almost degenerate in mass.

A comparison of these two flavor scenarios suggests that the flavor constraints on the
$U_{u,L}^{ij} = \delta^{ij}$ scenario are weaker and a substantial splitting in the Higgs
masses is allowed. With sizable splittings between the Higgs masses $M_{H^0}$ and 
$M_{A^0}$,  the electroweak precision tests are consistent with a lighter Standard 
Model-like Higgs, as is shown in Fig.~\ref{fig:DD}.  For the $U_{u,L}^{ij} = 
V_{\rm CKM}^{*ji}$ scenario, instead, the flavor constraints typically require
a small splitting between the $H^0$ and $A^0$ Higgs masses and hence 
a heavier Standard  Model-like Higgs boson is preferred in this case.

\section{$W^{+}H^{0}/W^{+}A^{0}$ production and W+jets Excess}
The signal process of interest to us is 
\[
p\bar{p}\to H^{\pm}\to W^{\pm}H^{0}/W^{\pm}A^{0}\to\ell^{\pm}\nu jj\,,
\]
where the charged lepton comes from the $W$-boson decay. 
This is  the dominant source of $W^\pm H^0/A^0$ production since the charged Higgs is produced resonantly. In addition there are t-channel processes that contribute to the total signal rate. Inverting the hierarchy between the neutral and charged Higgs boson, one could also consider the process 
\[
p\bar{p}\to H^{0}/A^{0}\to W^{\pm}H^{\mp}\to\ell^{\pm}\nu jj\,.
\]
This will have essentially a very similar collider signature and comparable effects in flavor and electroweak measurements.

In addition to the desired signal, our model also contributes to the $Z+jj$ and $\gamma+jj$ channels, through the processes
\begin{align*}
& p\bar{p}\to ZH^{0}/A^0\to Zjj,\qquad p\bar{p}\to A^{0}\to ZH^{0} \to Zjj\,,\\
& p\bar{p} \to \gamma H^0/A^0 \to \gamma jj \,.
\end{align*}
Since these processes are not mediated by on-shell resonances, they are suppressed compared to the $W^\pm jj$ channel. The tree-level cross sections for these channels at the Tevatron are listed in Tab.~\ref{tab:cx}. 

The mass of the scalar $H^0$ is fixed to $150$~GeV in order to reproduce the excess in the dijet invariant mass spectrum. 
We calculate the cross sections for $W^\pm H^0/A^0$, $Z H^0/A^0$ and $\gamma H^0/A^0$ at the Tevatron for two values of the charged Higgs mass, $M_{H^\pm}=250$~GeV and $M_{H^\pm}=300$~GeV. Larger splittings between the neutral and charged Higgs states are disfavored by EWPTs. The remaining free parameter is the up quark Yukawa coupling $y_{u,2}^{11}$, which is chosen such that the required amount of signal events is obtained.  As discussed in the previous section, we shall assume that both $y_{u,2}^{21}$ and
$y_{u,2}^{31}$ are smaller than $y_{u,2}^{11}$ and hence they have no impact in the particle production at the Tevatron. We shall briefly discuss on the possible effects of a different coupling arrangement in the conclusions of this article. 

\begin{table}[t]
\begin{center}
\begin{tabular}{|c||c|c|c|c|c|}
\hline 
$M_{H^\pm}$ [GeV], $y_{u,2}^{11} $  & $\Gamma_{H^\pm} $ [GeV] & Br$(H^\pm \to W^\pm \varphi^0)$ & $\sigma (W^\pm \varphi^0) $ &$ \sigma(Z \varphi^0)$ & $\sigma(\gamma \varphi^0)$  \\\hline \hline
300, 0.06 & 4.45 & 97.4\% & 2.04 & 0.12 & 0.11  \\\hline
250, 0.06 &  0.587 & 84 \% & 3.74 &  0.12 & 0.11 \\\hline
\end{tabular}
\end{center}
\caption{\small Branching fractions and production cross sections for two benchmark points. The remaining parameters are $M_{H^0} = M_{A^0} = 150$~GeV. 
 We display the total width of $H^\pm$ and the branching fraction into $W^\pm \varphi^0$, where $\varphi^0 = \{H^0,A^0\}$. The last three columns show the cross sections $\sigma(p \bar{p} \to X \varphi^0)$ in pb at the Tevatron. For the photon $p_T > 20$~GeV is required. 
}
\label{tab:cx}
\end{table}%

The signal is generated using CalcHEP/CompHEP~\cite{Pukhov:2004ca,Boos:2004kh} (and checked independently with MadGraph~\cite{Alwall:2007st}), and then processed through Pythia~\cite{Sjostrand:2006za} for parton showering and hadronization and through PGS for a detector simulation, for which the CDF parameter set is used. 

We implement the cuts of the CDF analysis: for the leptons, we require $p_T > 20$~GeV and $|\eta|< 1$. In addition, a missing energy greater than $30$~GeV is required, and the transverse mass of the lepton+$E\!\!\!/_T$ system is required to be larger than $30$~GeV. Jets are reconstructed with a cone size $R=0.4$ and required to have $p_T>30$~GeV, $|\eta|<2.4$ and $\Delta \eta_{jj} < 2.5$. Further the transverse momentum of the dijet system is required to have $|p_{Tj_1} + p_{Tj_2}|>40$~GeV. 
Events with more than two jets are rejected, as well as events with an additional lepton with $p_T>20$~GeV. We also reject events where the lepton is within a $\Delta R \leq 0.52$ cone around either of the two jets.

To correct the jet energy scale, we also simulate the $W+{\rm 2jets}$ signal from $W^+W^-$ and $W^\pm Z$ production. 
Then we multiply the jet energy by $1.1$ such that the mass of the hadronically decaying $W$-boson is reconstructed at $80$~GeV. The same scaling is applied to the signal. 

\begin{figure}[t]
\center
\includegraphics{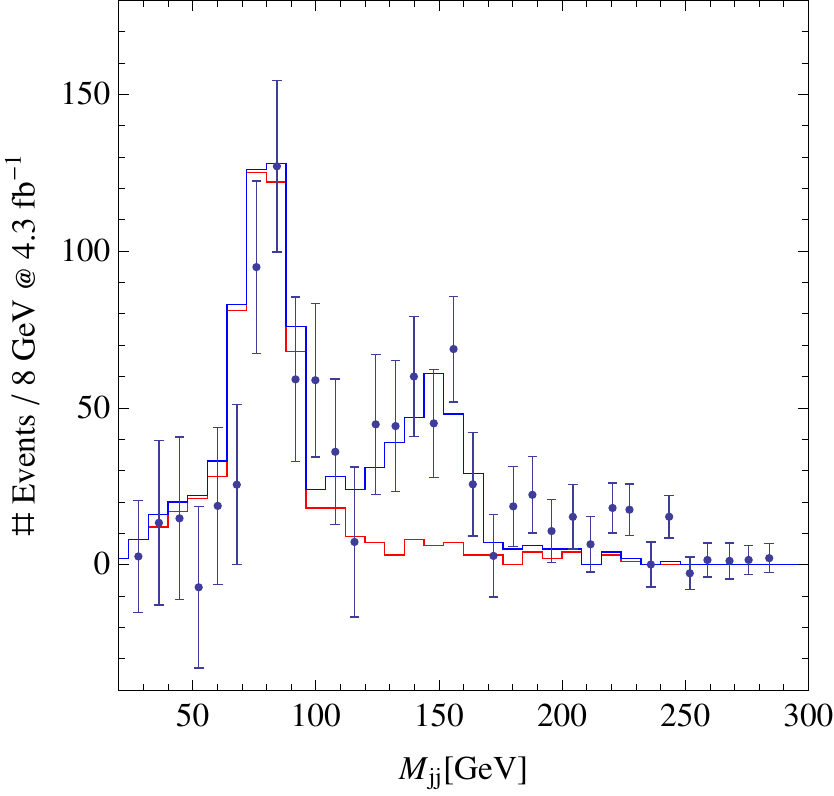}
\caption{Dijet invariant mass spectrum for di-boson background only (red) and signal plus background (blue). The parameters chosen for the simulation are $M_{H^\pm}=300$~GeV, $M_H^0 = M_A^0 = 150$~GeV and $y^{11}_{u,2}=0.06$. The data points are taken from Fig.~1 in~\cite{Aaltonen:2011mk}.
}
\label{fig:sim}
\end{figure}

The result of the simulation for di-boson background only and for signal plus di-boson background is shown in Fig.~\ref{fig:sim}. To match the height of the W-boson peak to data we multiply the tree-level cross sections for $WW$ and $WZ$ production by a K-factor of 1.4. 
For $M_{H^\pm} = 300$~GeV, 17\% of the events with $W\to \ell \nu_\ell$, $(\ell = e,\mu)$ pass the cuts, while for $M_{H^\pm} = 250$~GeV this number drops to 8.7\%. 
A good fit to the dijet invariant mass spectrum is found using 300 signal events, corresponding to a coupling $y_{u,2}^{11} = 0.06$ for both $M_{H^\pm}=300$~GeV and $M_{H^\pm}=250$~GeV. 

In Ref.~\cite{Eichten:2011sh} a different set of cuts was suggested for the CDF analysis that may improve the signal to background ratio by a factor of two. In particular, to be more sensitive to the $W$+resonance structure of the event, they suggest a cut on the sum of lepton and missing transverse momentum,  $|p_{T\ell}+p_{T\nu_\ell}|>60$~GeV, and require the angular separation between the jets to be $\Delta\phi(j_1,j_2)>1.75$.  For our model with $M_{H^\pm}=300$~GeV, these cuts reduce the signal by 32\%. Comparing with the results of Ref.~\cite{Eichten:2011sh}  we expect a similar increase of the signal to background ratio in our case.

Let us finally comment on the effects of splitting the masses of the neutral Higgs bosons $H^0$ and $A^0$.  Small mass differences have only a minor influence on the
signal rate. For large mass differences, $\delta m_{H^0A^0} \geq$ 50 GeV,  as required for a significant improvement  of the fit to  precision electroweak data for  moderate values of the SM Higgs boson mass, 
the dominant signal rate is associated with the lightest of these Higgs bosons, while the other one leads to a small or no signal contribution. This is, for instance, the case for the values $M_{H^0} \simeq 150$~GeV,
$M_{A^0} \simeq 210$~GeV, and  $M_{H^\pm} \simeq 300$~GeV,  used in Fig.~\ref{fig:EWPT}.  Although the charged Higgs may decay to on-shell $W$ and $A^0$ bosons,   the phase space suppression is strong enough to lead to only a very minor increase of the signal in the $M_{jj} \simeq 210$~GeV region. This region can  however be populated if also the mass of the charged Higgs is increased (see Appendix). 
\section{Prospects at Tevatron and LHC}
At hadron colliders, the $150$~GeV dijet resonance can be produced also in association with a $Z$ boson or with a photon. In our model, the main production of the $W+{\rm jets}$ signal is through resonant charged Higgs production. Hence, the $\gamma+{\rm jets}$ and $Z+{\rm jets}$ signals at the Tevatron, which are only produced in the t-channel, are suppressed by more than an order of magnitude with respect to the $W +$ 2jets one (c.f. Tab.~\ref{tab:cx}). Assuming a similar signal acceptance, and taking into account the small branching fraction of the $Z$ boson to lepton pairs, we do not expect the Tevatron to see any relevant excess in the $(Z\to \ell \ell)jj$ channel~\cite{talk1}.

In the photon plus dijet channel at the Tevatron, we expect up to 40 events for  a photon with $p_T>20$~GeV. This channel is dominated by QCD dijet production with an additional photon. A search in this channel has been performed with very mild cuts on the jet energies~\cite{CDFnote:10355}, and our model is consistent with these observations.  A more refined analysis with optimized cuts could provide important constraints on this and other models for the $W$+jets CDF excess.

At the LHC, it is possible to search for all of the above signals. 
In addition, the larger center of mass energy allows for sizable production cross sections also for $ZH^\pm$, $W^\pm H^\mp$ and for pair production of scalars.  The cross sections at the LHC with $\sqrt{s}=7$~TeV are given in Tab.~\ref{tab:LHC}. 

\begin{table}[t]
\begin{center}
\begin{tabular}{|c||c|c|c|c|c|c|}
\hline 
$M_{H^\pm}$ [GeV], $y_{u,2}^{11} \quad$ & $\sigma (W^\pm \varphi^0) $ &$ \sigma(Z \varphi^0)$ & $\sigma(\gamma \varphi^0)$  & $\sigma(W^\pm H^\mp)$& $\sigma(Z H^\pm)$ &  $\sigma(H^+H^-)$ \\\hline \hline
300, 0.06 & 7.72 & 0.35 & 0.31 & 0.026 & 0.011 &  0.0010 \\\hline
250, 0.06 & 13.7 &  0.35 & 0.31& 0.034 & 0.016 &  0.0023 \\\hline
\end{tabular}
\end{center}
\caption{\small LHC production cross sections for two benchmark points. 
The remaining parameters are $M_{H^0} = M_{A^0} = 150$~GeV. Shown are the production cross sections (in pb) at LHC with $\sqrt{s}=7$~TeV for several promising channels. For the photon $p_T > 20$~GeV is required. 
}
\label{tab:LHC}
\end{table}%

One promising signal comes from Drell-Yan production of $H^+ H^-$ pairs, that gives rise to $W^+ W^-$+4jets signals. The main background for this process is $t\bar{t}$ production with additional jets. This background can be reduced by explicitly asking for four hard jets, and demanding that the invariant mass of pairs of jets is close to the mass of the resonance. In addition, a veto on b tagged jets may be useful to discriminate between the signal and background.

\section{Conclusions} 
We have presented a simple renormalizable model that can explain the CDF excess in $W$+2jets that was recently observed by the CDF collaboration.  Our model is in agreement with all constraints coming from electroweak precision measurements, and  depending on the precise values of the new charged and neutral Higgs boson masses, allows  a broad range of  the Standard Model Higgs mass from 100 GeV to several hundreds of GeV.

The model predicts no significant signal in the $Z$+2jets and in the $W b\bar{b}$ channel, in agreement with current experimental constraints on these channels. At the LHC, the model can be searched for in dijets produced in association with an electroweak gauge boson. In addition, we also expect a sizable production cross section for $WW$+jets and $ZW$+jets at the LHC. These signals might be easier to separate from the notoriously large QCD background. 

Given the steadily growing number of models that attempt to explain the CDF excess, it might be worth considering possibilities to discriminate between different models. In principle angular observables should be suitable for this task, since angular distributions typically depend on the spin of intermediate resonances and on the production mode. 

More concretely, we expect t-channel $Z'$ models to give a different $p_T$ spectrum (softer and more forward) than in the s-channel production considered in this work.  Moreover, no resonant feature on the total invariant mass of the $W + jj$ system is expected in those models. Considering there is only one leptonic $W$, we can fully reconstruct the event and study the spin of various resonances. In our particular case, the spin of $H^+$ may be studied using the $W H^0$ angular distribution in the rest frame of $H^+$.  As stressed before, the relative ratio of $W + jj$ rate to the $Z + jj$ (or $\gamma$ + jj) is a very good diagnostic that may distinguish different models. 

We finally want to mention that, had we assumed $y_{u,2}^{31}$ to be the most significant coupling, a relevant signal could still be obtained for large values
of this coupling, $y_{u,2}^{31} \simeq 1.5$, via the $u \bar{b} \to H^+ \to H^0/A^0 + jj$ and $g u \to t H^0/A^0$ channels. We studied this possibility and found 
that the rate in the former channel is a factor four larger than the latter and therefore the signal characteristics are very similar to the one studied in this article.  
Similarly to what happens in the model of Ref.~\cite{Nelson:2011us},  such a coupling arrangement leads also to an increase of the top-quark forward-backward asymmetry, 
but is subject to constraints coming from flavor physics and top-quark decays~\cite{Zhu:2011ww}.  Further discussion on these possibilities will be studied elsewhere.%

~\\
{\bf Acknowledgements} \\
 
We thank W.~Altmannshofer and R.~Culbertson for useful discussions. 
Work at ANL is supported in part by the U.S. Department of Energy (DOE), Div. of HEP,
Contract DE-AC02-06CH11357. Fermilab is operated by the Fermi Research Alliance, LLC under Contract No 
DE-AC02-07CH11359 with the U.S. Department of Energy.
 P.S. is partially supported by the UIC DOE HEP
Contract DE-FG02-84ER40173. L.-T.W. is supported by the DOE Early
Career Award under grant DE-SC0003930. A.M. is supported by the U.S. Department of Energy under Contract No. DE-FG02-94ER40840.

\appendix

\section*{Appendix}

The mass spectrum of the model is strongly constrained by precision constraints. Mass splittings between the scalars larger than the ones considered above are very hard to bring in agreement with these constraints. However, one could also view this model as an effective theory valid up to around the TeV scale, where additional new physics is expected to solve the hierarchy problem. In this case, additional new physics contributions to the electroweak S and T parameters will in general be present. These may partially compensate the contributions from the Higgs doublet, thus making larger mass splittings phenomenologically viable. 

\begin{figure}[t]
\center
\includegraphics{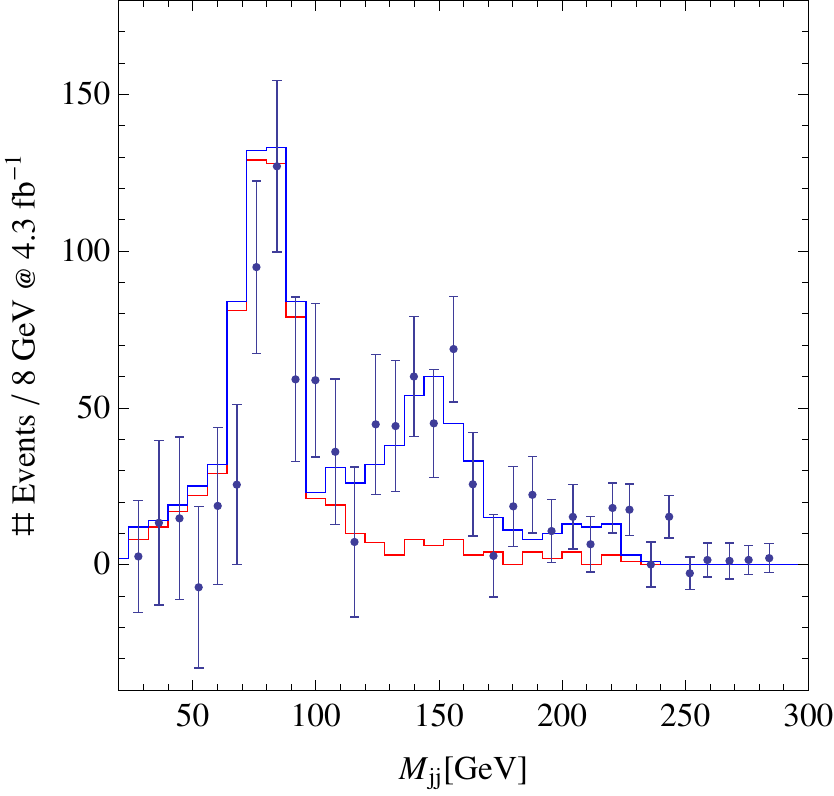}
\caption{Dijet invariant mass spectrum for di-boson background only (red) and signal plus background (blue). The parameters chosen for the simulation are $M_{H^\pm}=330$~GeV, $M_{A^0} = 210$~GeV and  $M_{H^0} = 150$~GeV. 
}
\label{fig:fitall}
\end{figure}

Although the data in the invariant mass bins between 180~GeV and 240~GeV lie systematically above the expected background, this excess is not statistically  significant, and in addition, is subject to possibly large NLO uncertainties~\cite{talk}.  Consequently, no preference for a model leading to such an excess can be claimed, on the other hand, the data is not in conflict with this possibility.  As an example of such a model, in Fig.~\ref{fig:fitall} we show the dijet invariant mass distribution for $M_{H^\pm}=330$~GeV, $M_A=210$~GeV and $M_{H^0}=150$~GeV. In this case our model fits well the data  up to invariant masses of $230$~GeV, while only a small increase of the coupling $y_{u,2}$ is required.

\end{document}